\documentclass[journal]{IEEEtran}
\usepackage{cite}
\usepackage{graphicx}
\graphicspath{{figs/}}
\usepackage{stfloats}
\usepackage{amsmath}
\usepackage{amssymb}
\usepackage{array}
\usepackage{amsfonts}
\usepackage[normalem]{ulem}
 
\usepackage[utf8]{inputenc}
\usepackage{tikz}
\usetikzlibrary{positioning}
\usetikzlibrary{shapes.geometric, arrows}
\usepackage{xcolor}

\usepackage{booktabs}
\usepackage{algorithm}
\usepackage{ifthen}
\usepackage{caption}
\usepackage{subcaption}
\usepackage[noend]{algpseudocode}
\MakeRobust{\Call}
\usepackage{url}
\newcommand{\R}{\mathbb{R}}

\title{Training Variational Networks with Multi-Domain Simulations: Speed-of-Sound Image Reconstruction}

\author{Melanie~Bernhardt,\ \ Valery~Vishnevskiy,\ \ Richard~Rau,\ \ Orcun~Goksel\\[1ex]
        \textit{Computer-assisted Applications in Medicine, ETH Zurich, Switzerland}
\thanks{
Funding was provided by the Swiss National Science Foundation and Innosuisse.
Corresponding author is O.~Goksel (ogoksel@ethz.ch).}}

\begin{document}
\maketitle
\begin{abstract}
Speed-of-sound has been shown as a potential biomarker for breast cancer imaging, successfully differentiating malignant tumors from benign ones. 
Speed-of-sound images can be reconstructed from time-of-flight measurements from ultrasound images acquired using conventional handheld ultrasound transducers. 
Variational Networks (VN) have recently been shown to be a potential learning-based approach for optimizing inverse problems in image reconstruction.  
Despite earlier promising results, these methods however do not generalize well from simulated to acquired data, due to the domain shift. 
In this work, we present for the first time a VN solution for a pulse-echo SoS image reconstruction problem using diverging waves with conventional transducers and single-sided tissue access.
This is made possible by incorporating simulations with varying complexity into training. 
We use loop unrolling of gradient descent with momentum, with an exponentially weighted loss of outputs at each unrolled iteration in order to regularize training.
We learn norms as activation functions regularized to have smooth forms for robustness to input distribution variations. 
We evaluate reconstruction quality on ray-based and full-wave simulations as well as on tissue-mimicking phantom data, in comparison to a classical iterative (L-BFGS) optimization of this image reconstruction problem. 
We show that the proposed regularization techniques combined with multi-source domain training yield substantial improvements in the domain adaptation capabilities of VN, reducing median RMSE by 54\% on a wave-based simulation dataset compared to the baseline VN. 
We also show that on data acquired from a tissue-mimicking breast phantom the proposed VN provides improved reconstruction in 12 milliseconds.
\end{abstract}

\section{Introduction}

Quantitative ultrasound (US) biomarkers are essential for several clinical applications for diagnosis and staging.  
With ultrasound computed tomography (USCT), tissue properties such as speed-of-sound (SoS) and attenuation can be characterized~\cite{duric_detection_2007,gemmeke_3d_2007,Li2010,mamou_quantitative_2013,malik_breast_2019}. 
SoS as an imaging biomarker may have clinical applications such as breast tissue classification~\cite{klock_visual_2017,sanabria_breast-density_2018}, solid mass differentiation~\cite{iuanow_accuracy_2017,LisaRuby-InvestigativeRadiology-Jul2019}, quantifying muscle loss~\cite{sanabria_speed_2018_sarc}, and imaging human-knee~\cite{wiskin_3d_2019}.
Typical USCT setups operate in transmission-mode on tissue suspended in a water bath, e.g., with opposing~\cite{malik_quantitative_2018}, ring-shaped~\cite{duric_detection_2007}, or full 3D~\cite{gemmeke_3d_2007} transducer geometries.
These however are bulky and costly setups, which can image only submersible anatomical structures. 

Recently, several SoS imaging methods have been proposed using conventional US systems with hand-held probes. 
In~\cite{Krueger98,Huang2004,SanabriaMICCAI,Sanabria:SoSReflector} a passive acoustic reflector was used to allow recording time-of-flights (ToF) of reflected signals. 
SoS distribution can then be reconstructed by solving an ill-posed limited angle inverse problem.
Such a reflector-based method was recently extended for attenuation mapping in~\cite{rau_attenuation_2019}.
In~\cite{jaeger_computed_2015} echos from tissue scattering were used to obtain relative ToF readings by insonifying from multiple plane wave angles, allowing SoS reconstruction using a Fourier-domain inversion, with a direct solution approach and in-vivo images presented in~\cite{JAEGER2015}.
A spatial-domain regularized inverse-problem was shown in~\cite{Sanabria:spatialreconstsos} to yield improved results, shown successful for differential diagnosis of breast cancer in~\cite{LisaRuby-InvestigativeRadiology-Jul2019}.
SoS reconstructions were shown in~\cite{jaeger_full_2015,rau_aberration_2019} to help correct beamforming time-delays, hence improving B-mode resolution.
Adapting receive beamforming apertures to achieve invariant PSF was shown to facilitate reconstructions in~\cite{stahli_forward_2019}.
Recently diverging waves acquisition was proposed in~\cite{rau_sos19} to yield superior reconstructions thanks to reduced wavefront aberrations compared to plane waves. 

Solving the ill-posed reconstruction problem to obtain such SoS maps is a challenging task that requires carefully chosen regularization and numerical optimization techniques. 
These difficulties have motivated the introduction of machine learning based methods to find the best regularization parameters. 
In \cite{Zhu18}, an end-to-end structure was utilized for learning inverse Fourier transform in MRI image reconstruction. A similar approach was utilized in SoS image reconstruction in~\cite{Feigin:sosendtoend}.
Variational Networks (VN) was introduced in~\cite{hammernik:vnMRI}, in which each network layer models one unrolled gradient descent step of optimizing the reconstruction inverse problem. 
In~\cite{Vishnevskiy:ReconstructVN}, VNs show promising results for solving SoS image reconstruction problem, when trained in a supervised manner using simple simulations via forward-problem with ray-tracing. 
However, supervision from such inaccurate synthetic data rarely translate successfully to phantom or in-vivo reconstructions, due to the large domain shift.  
A supervised learning from in-vivo data is not feasible since ground truth in-vivo SoS distributions are not known. 
More complex wave simulations are possible, which are however prohibitive for large training sets and also may overfit VN to any inaccuracies in such simulation pipeline.
Accordingly, we herein propose a VN approach using training data from multiple simulation setups and domains, in order to increase robustness to domain shift in order to enable SoS reconstructions in real acquired data, without ever encountering one in the training stage. 

\textit{Domain shift} is the problem occuring when the training data distribution differs from the distribution of testing data, and \textit{domain adaptation} is the group of approaches to successfully transfer the learned information to the test samples despite a domain shift~\cite{kouw:domainadapt}. 
With several source domains for training, the domain adaptation problem is known as \textit{multi-source domain adaptation}~\cite{Sun}. 
Several approaches have been proposed for combining information from different domains, such as feature representation to combine domains in training, learning a data-dependent regularizer~\cite{Duan:domain,Duan:video}, and weighting domain sources based on their probability~\cite{Chatto}; with an extended review provided in~\cite{Sun}.
\emph{Multi-task learning} has a similar motivation, where the information from multiple related tasks is believed to help improve the generalization performance of learned models~\cite{Zhang:MTL}. 

In this work, we focus on solving the inverse-problem of pulse-echo sound-speed imaging, using variational neural networks representing the unrolled iterations of a conventional optimization algorithm, where typically heuristic setting of regularizers, filters, and preconditioning schemes are learned from simulated data. To achieve this goal, we employ three major approaches to make VN-based reconstruction robust to domain shift from simulations to acquired data.
First, we adopt an exponentially weighted loss function integrating intermediate unrolled VN iterations~\cite{vishnevskiy:ct}, which helps to regularize the VN training and greatly stabilize the learned VN behaviour.
Second, we propose to regularize activation function forms that are used for learning arbitrary norms in the VN, which facilitates more easily generalizable and transferable activation functions.
In addition to the two VN adaptations above, last and most importantly, we utilize simulations with different models and levels of complexity to adapt the VN to different input distributions as well as best utilize the simulation computation time. Moreover, the proposed approach is by design able to handle varying amounts of missing measurements, i.e.\ excluded time-of-flight data from inaccurate displacement estimation.
Utilization of simulations from different models and complexity, enabled by the methods introduced herein and listed above, are demonstrated for the first time in this work, with an application on SoS image reconstruction in pulse-echo mode, using diverging waves with conventional linear US transducers.
Our results demonstrate that such multi-modal training together with our proposed adaptations yields major improvements in the generalization ability of VN, by  exploiting the diversity of different simulation models.

\section{Pulse-Echo SoS Imaging}
\begin{figure}
    \centering
    \includegraphics{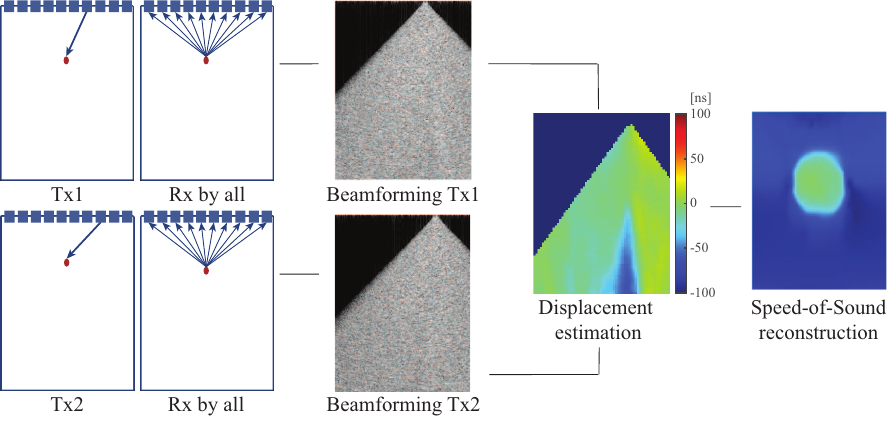}
    \caption{Illustration of SoS reconstruction pipeline using diverging waves~\cite{rau_sos19} from two single element transmissions. }
    \label{fig:fig1}
\end{figure}

Local SoS can be reconstructed from ToF deviations of US wavefronts~\cite{jaeger_computed_2015,Sanabria:spatialreconstsos,rau_sos19}. 
By approximating the wavefront propagation with straight lines on a discretized field of view of $P_x$ cells, integral ToF deviations $\mathbf{d} \in \R^{n_{r}}$ for different wave propagation paths can be related to local SoS by~\cite{Sanabria:spatialreconstsos}
\begin{equation}
\label{eq:forward}
    \mathbf{L}\mathbf{x}=\mathbf{d},
\end{equation}
where $\mathbf{x}\in\R^{P_x}$ is a vector of \emph{slowness} (inverse of SoS) values, $\mathbf{L}\in\R^{n_{r}\times P_x}$ describes the forward problem (FP) geometrically with the differential lengths $ l_{p,c} $ of path $p$ within cell $c$. 
Differential ToF measurements, $\mathbf{d}$, are obtained by measuring apparent displacements between different US propagation paths, which are then expressed geometrically in the FP with the $\mathbf{L}$ matrix.
We herein utilize an acquisition with diverging waves~\cite{rau_sos19}, with its corresponding FP and displacement estimations.
Acquisition is then accomplished by firing different transducer elements separately, and then recording for each the echo with all transducer elements, i.e.\ full-matrix capture (FMC). 
For each single-element diverging-wave transmit event, the received echo data from all receive elements is delay-and-sum beamformed to generate a spatial RF frame. Differential ToF measurements are obtained using local displacement tracking between pairs of such beamformed frames, from transmits of nearby elements as illustrated in Figure~\ref{fig:fig1}. 

Having acquired these differential ToF measurements, it is possible to reconstruct local SoS by defining a regularized objective function based on the FP introduced in Eq.~(\ref{eq:forward}). 
In~\cite{SanabriaMICCAI,Sanabria:SoSReflector} \textit{anistropically-weighted total variation} (AWTV) was shown to be a successful regularization strategy for a reflector based SoS reconstruction, in order to cope with the limited angle nature of this reconstruction problem. 
The inverse problem (IP) can then be solved using the following optimization formulation:
\begin{equation}  
\label{eq:IP}
\mathbf{\hat{x}} = \arg\min_{\mathbf{x}} \|\mathbf{L}\mathbf{x} - \mathbf{d}\|_1 + \lambda \|\mathbf{D}\mathbf{x}\|_\text{AWTV},
\end{equation}
where $\mathbf{D}$ is a regularization matrix and $\lambda$ the regularization weight.
In \cite{Sanabria:spatialreconstsos}, this regularizer was applied for spatial pulse-echo SoS reconstruction and was extended to a \emph{multi-angle} setting defined as $\|\mathbf{D}\mathbf{x}\|_\text{MA-AWTV} = \sum_{i,j}\sum_{\theta}\kappa_{\theta}|\mathbf{D}_{\theta}\mathbf{x}|$,
where $\mathbf{D}_{\theta}$ directional derivative along the unit vector with angle $\theta$. In~\cite{rau_sos19}, the regularization was further refined by using Sobel and Roberts kernels to regularize the spatial gradients.  

Having desired regularization forms, the above optimization problem can be then solved using traditional optimizers such as gradient descent or a quasi-Newton method e.g.\ limited-memory Broyden–Fletcher–Goldfarb–Shanno (L-BFGS) algorithm~\cite{broyden:bfgs,Fletcher:bfgs,Goldfarb:BFGS,Shanno:bfgs}.
Given memory and computational efficiency, convergence properties, and earlier successful utilization in SoS reconstruction~\cite{rau_sos19,Vishnevskiy:ReconstructVN}, we herein use L-BFGS as our baseline classical solver. 

\section{Variational Networks for Reconstruction}
For successful image reconstruction in general as well as in this particular limited-angle computed tomography setting, determining a suitable form of the IP as well as its parametrization is essential.
For instance in~\eqref{eq:IP}, the regularization form, weight, and norms of data and regularization terms, as well as iterative optimization parameters such as step-lengths play a large role in the robustness and accuracy of such solutions.
Indeed, these regularization parameters are difficult to set manually, as exemplified by the research works in the earlier subsection as well as in similar iterative optimization problems such as on elastography and registration.
Suitable optimization settings often require many assumptions, prior experience, and trial-and-error, without being able to guarantee whether some optimal settings are found eventually. 
This motivates VN for establishing a generic iterative optimization setting, where the IP definition and parametrization can be learned, while keeping the FP fixed to that known from physics and acquisition sequence~\cite{Vishnevskiy:ReconstructVN}. 

\subsection{Variational networks for pulse-echo SoS reconstruction} 
A generic regularized IP can be defined as
\begin{equation}
 \hat{\mathbf{x}} = \arg \min_{\mathbf{x}} \|\mathbf{L}\mathbf{x} - \mathbf{d}\| + \lambda\mathcal{R}(\mathbf{x}),  
\end{equation}
which aims to find the maximum-a-posteriori (MAP) solution for a FP as in~\eqref{eq:forward},
where $\mathcal{R}$ denotes a generic regularization function. 
A simple gradient-descent solution of such IP would then iteratively update an estimate of $\mathbf{x}$ as follows:
\begin{equation} \label{eq:iter}
    \mathbf{x}^{(i)}  = \mathbf{x}^{(i-1)} - \eta^{(i)} \cdot \left [ \mathbf{L}^T(\mathbf{L} \mathbf{x}^{(i-1)} - \mathbf{d}) + \lambda \nabla_{\mathbf{x}} \mathcal{R}(\mathbf{x}^{(i-1)}) \right],
\end{equation}
where $i$ is the iteration and $\eta^{(i)} \in \mathbb{R}$ is the step size.

In~\cite{hammernik:vnMRI}, it was proposed to unroll the iteration loops of such gradient descent algorithm as network layers, which perform the mathematical operations indicated in~\eqref{eq:iter} with $i$ being a network layer index.
It was successfully applied for MRI reconstruction in~\cite{hammernik:vnMRI}, where the gradient of the regularizer was further parameterized by convolution kernels and non-linear potential functions as follows:
\begin{equation}  \label{eq:R}
    \nabla\mathcal{R}(\mathbf{x}^{(i)}) =   \sum_{j=1}^{N_i} \mathbf{K}_j^{(i)T} \mathbf{\phi}^{(i)}_j(\mathbf{K}_j^{(i)} \mathbf{x}^{(i-1)}),
\end{equation}
with $\mathbf{K}_j^{(i)}$ denoting convolution kernels (i.e. filters), $N_i$ the number of convolution of kernels for layer $i$, $\phi^{(i)}_j$ non-linear potential functions. 
This regularization scheme with multiple regularization terms is known as the \textit{Field of Experts} model. 

In such VN representing loop-unrolled iterations, the parameters that would normally be manually set can then be learned with supervision from sample ground truth reconstruction pairs in a deep learning setting.
For instance, the set of parameters to learn in~\eqref{eq:R} would be $ \{ \mathbf{\phi}^{(i)}_j, \{\mathbf{K}_j^{(i)}\}_{j=1}^{N_i}, \mathbf{\eta}_i \}_{i=1}^{I},$ where $I$ is the number of unrolled iterations. 
To initialize loop-unrolling, we herein used $\mathbf{x}^{(0)} = \mathbf{L}^T \mathbf{d}$ as a rough estimate of SoS reconstruction, similarly to~\cite{hammernik:vnMRI,Vishnevskiy:ReconstructVN,vishnevskiy:ct}. 
The desired reconstruction is then the final VN output from the last unrolled update step, i.e.\ $\mathbf{\hat{x}}:=\mathbf{x}^{(I)}$. 

\begin{figure}
    \centering
    \includegraphics[width=\linewidth]{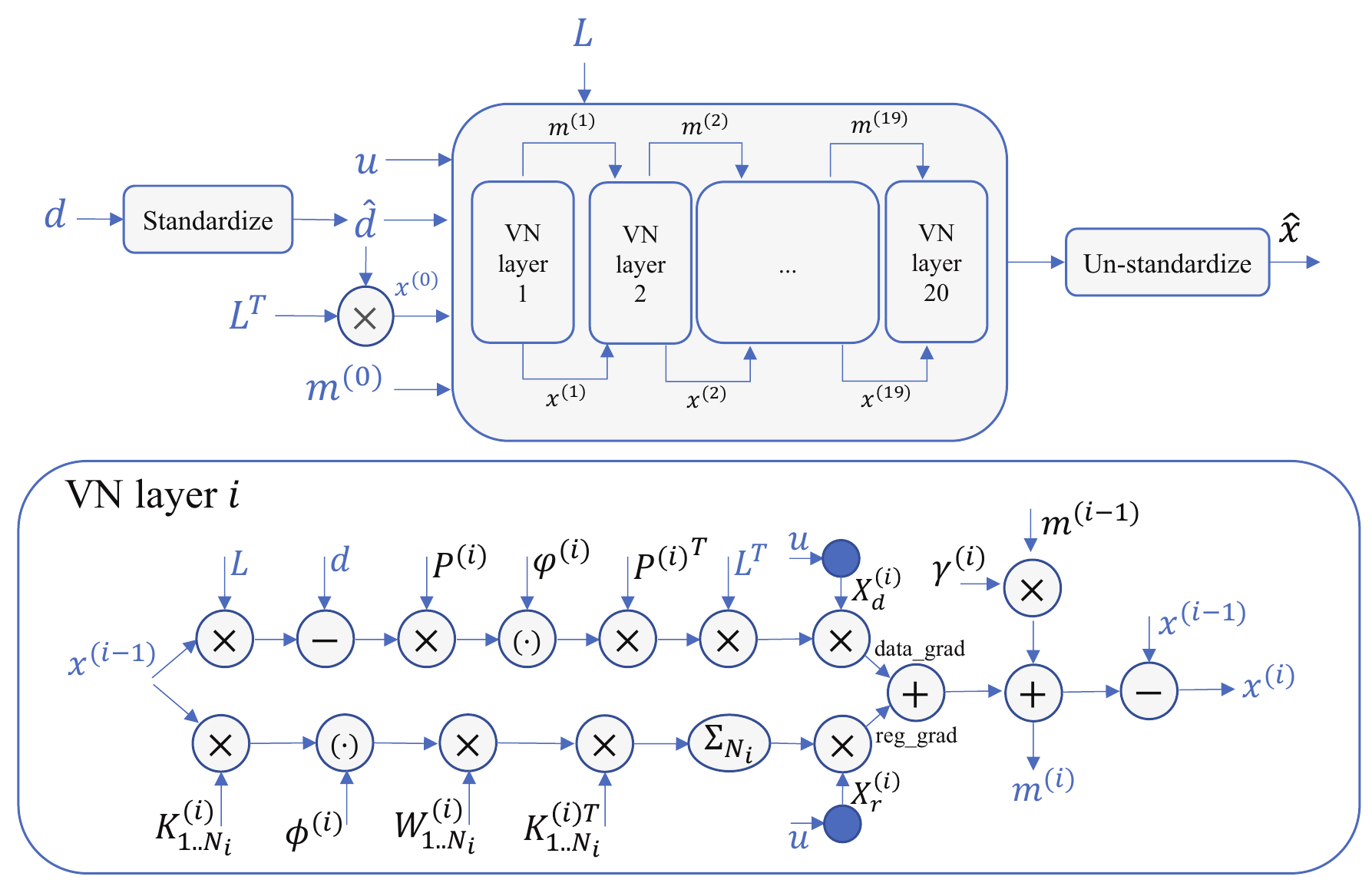}
    \caption{Overview of the architecture of the proposed network. Inputs and outputs are in blue, trainable parameters are in black, and ``*'' and ``$\times$'' represent convolution and matrix multiplication.}
    \label{fig:fig_network}
\end{figure}

Our overall VN structure is summarized with the forward pass prototype in Algorithm~\ref{algo:finalSoSVN} and is depicted schematically in Fig.~\ref{fig:fig_network}. Note that the $L$ matrix is constructed once in an initialization step following \cite{rau_attenuation_2019} based on the employed probe geometry, image sequence, beamforming, and displacement tracking parameters and their estimated physical consequences. Accordingly, the parameters need to be learned separately if different acquisition settings are desired.
In this work, we adopt loop unrolling of gradient descent with momentum~\cite{qian:momentum}, as it was proposed in~\cite{Vishnevskiy:ReconstructVN} to have more robust convergence. 
We use spatial filter weights $\mathbf{W}_j^{(i)}$ in~\eqref{eq:R} to allow for different filters to be active over different regions of the images~\cite{Vishnevskiy:ReconstructVN} and allowing to approximate the gradient of various regularizers. 
We also learn a diagonal left preconditioner $\mathbf{P}^{(i)}$ such that $\mathbf{P}^{(i)}\mathbf{L}$ is better conditioned than $\mathbf{L}$, as shown in~\cite{vishnevskiy:ct} to be essential for ill-conditioned problems such as limited-angle CT.
Furthermore, inspired by the potential functions $\phi$ in~\eqref{eq:R}, an adaptive data fidelity term is also learned via a non-linear potential function $\mathbf{\psi^{(i)}}$ for the data term, cf.\ line~5 in Alg.~\ref{algo:finalSoSVN}. Table \ref{tab:1} summarizes all the learned parameters.

Note that in practice data from actual acquisitions, such as the estimated displacements $\mathbf{d}$ herein, would often contain noise. 
Such noisy measurement components can sometimes be predicted to some extent given some form of confidence measure, e.g. the displacement estimation correlation coefficient in our setting.
Since highly noisy measurements would be detrimental in the IP, irrespective of the regularization form employed, one can filter out and omit very noisy measurements from the solution, i.e.\ remove the corresponding rows of $\mathbf{L}$ and $\mathbf{d}$ or set corresponding elements of $\mathbf{d}$ to zero (similarly to zero-masked undersampling in MRI).
Varying number of measurements left from such confidence masking (herein called \emph{undersampling rate}) would then affect the optimization problem and the parameterization thereof differently; for example smaller number of measurements potentially requiring a higher regularization weight.
Accordingly, we propose herein an undersampling rate dependent weighting of the regularizer and data term, by defining two non-linear 1D functions $\mathbf{\chi}_d^{(i)}$ and $\mathbf{\chi}_r^{(i)}$ of the undersampling rate $u$ for the data and the regularization terms, respectively, as seen on line~8 of Alg.~\ref{algo:finalSoSVN}. 

In order to increase the capacity of the network, instead of using standardized convolutional filters, we parameterize the mean of the convolutional filters. 
Accordingly, at each layer, the filters are first zero-centered and standardized, and then offset by adding a mean value learned for each filter separately, cf.\ line 7 in Alg.~\ref{algo:finalSoSVN}.  

\begin{algorithm}[t]
	\caption{Variational Network Forward Pass}
	\label{algo:finalSoSVN}
	\begin{algorithmic}[1]
		\Procedure{unroll}{L, d, u}
		\Statex $\triangleright \mathbf{L}$: ray-path matrix
		\Statex $\triangleright \mathbf{d}$: input measurement
		\Statex $\triangleright \mathbf{u}$: undersampling rate
		\Statex $\triangleright \mathbf{x}$: inverse SoS
		\State $\mathbf{d} \gets \Call{Standardize}{\mathbf{d}}$
		\State Initialize $\mathbf{x}_{0} \gets L^Td$, $\mathbf{m}^{(0)} \gets 0$ 
		\For{i = 1 to I = 20}
		\State data\_grad $\gets (\mathbf{P}^{(i)}\mathbf{L})^T\mathbf{\psi}^{(i)}\left \{\mathbf{P}^{(i)}(\mathbf{L}\mathbf{x}^{(i-1)} - \mathbf{d})\right \}$
		\State $\mathbf{K}_j^{(i)} \gets \frac{\mathbf{K}_j^{(i)}-\mathbf{\mu}_{\mathbf{K}_j^{(i)}}}{\sigma_{\mathbf{K}_j^{(i)}}}$  
		\State $\mathbf{K}_j^{(i)} \gets \mathbf{K}_j^{(i)} + \tilde{\mu}_{\mathbf{K}_j^{(i)}}$ 
		\State reg\_grad $\gets  \sum_{j=1}^{N_i}\mathbf{K}_j^{(i)T} \mathbf{W}_j^{(i)} \mathbf{\phi}^{(i)}_j\left \{\mathbf{K}_j^{(i)} \mathbf{x}^{(i-1)}\right \} $ \\
		\State total\_grad $\gets \mathbf{\chi}_d^{(i)}(u) \mbox{data\_grad} + \mathbf{\chi}_r^{(i)}(u) \mbox{reg\_grad}$ 
		\State $\mathbf{m}^{(i)} \gets \gamma^{(i)} \cdot \mathbf{m}^{(i-1)} + \mbox{total\_grad}$
		\State $\mathbf{x}^{(i)} \gets \mathbf{x}^{(i-1)} - \mathbf{m}^{(i)}$
		\EndFor
		\State $\hat{\mathbf{x}} \gets \Call{Un-standardize}{\mathbf{x}_I}$
		\State \Return{$\frac{1}{\hat{\mathbf{x}}}$}
		\EndProcedure
	\end{algorithmic}
\end{algorithm}
\begin{table}
\caption{Overview of the learned parameters, where [$\cdot$] indicate convolution kernel sizes and the linear interpolation functions are parameterized by their \emph{knots}.}
\label{tab:1}
\centering
\begin{tabular}{@{}>{\raggedright}p{0.10\linewidth}p{0.53\linewidth}p{0.23\linewidth}@{}}
\toprule
Parameter & Description & Dimension \\
\midrule
   $\mathbf{P}^{(i)}$  & Pre-conditioners & 64x84x6 per layer  \\
   $\mathbf{W}^{(i)}$ & Spatial weights & 57x77x32 \\
   $\mathbf{K}^{(i)}$ & Regularizers as convolutional filters & [8,8,32]\\
   $\tilde{\mu}_{K^{(i)}}$ & Mean values of convolutional filters & 32 \\
   $\phi^{(i)}$ & Regularization term activation & 35 knots per layer \mbox{+ 1 max value} \\
   $\psi^{(i)}$ & Data term activation & 35 knots\phantom{per layer} \mbox{+ 1 max value} \\
  $\chi^{(i)}_d $ & Undersampling weighting, data term & 35 knots \\
    $\chi^{(i)}_r $ & Undersampling weighting, regularization & 35 knots \\
   \bottomrule
\end{tabular}
\end{table}

\subsection{Standardization}
In tissues, SoS values vary around 1500\,m/s with the contrast being relatively small, up to 3-10\%.
The non-zero-centered nature of this problem thus negatively affects the residuals in optimization, especially considering the often zero-centered nature of typical activation functions in deep learning.
Accordingly, we propose to standardize the reconstructed SoS values by removing an offset equivalent to a homogeneous SoS background, which was indicated in our preliminary tests to be largely beneficial.
To determine the zero-centering offset, we first solve the following least squares problem representing a homogeneous-equivalent SoS reconstruction without any regularization
\begin{equation} \label{eq:k}
    k^\star := \arg\min_k \|k\mathbf{L}\mathbf{1}_{P_{x}}-\mathbf{d}\|_2^2 = \frac{\langle \mathbf{d}, \mathbf{L}\mathbf{1}_{P_{x}}\rangle}{\langle \mathbf{L}\mathbf{1}_{P_{x}}, \mathbf{L}\mathbf{1}_{P_{x}} \rangle},
\end{equation}
where $\mathbf{d}$ is the zero-masked undersampled measurements, $P_x$ is the dimension of $\mathbf{x}$, and $\mathbf{1}_{P_{x}}$ is a vector of ones with dimension $P_x$. 
This offset $k^\star$ is subtracted from the input measurements $d$, and then divided by a standard deviation approximate $s$ calculated as
\begin{equation}
    s^\star := \sqrt{\frac{\|\mathbf{d} - k^\star L \mathbf{1}_{P_{x}}\|_2^2}{\|\mathbf{d}\|_0}},
\end{equation}
where $\|\cdot\|_0$ denotes  the number of non-zero elements and is proportional to undersampling rate. 

Given above, with a change of variables the VN can then be used to solve
\begin{equation}
 \mathbf{L}\mathbf{x}' = \mathbf{d}'
 \iff \mathbf{L}\left (\frac{\mathbf{x}  - k^\star \cdot \mathbf{1}_{P_{x}}}{s^\star}\right) = \frac{\mathbf{d} - k^\star\cdot \mathbf{L}\mathbf{1}_{P_{x}}}{s^\star},
\end{equation}
where the prime variables indicate standardized forms of the measurements $\mathbf{d}$ and the inverse SoS values $\mathbf{x}$, which accordingly removes the contribution of a homogeneous SoS background field in order to better condition the optimization problem.
In practice for inference given measurements, we first solve the least-squares problem~\eqref{eq:k}, next standardize $\mathbf{d}$ and initialization $\mathbf{x}^{(0)}$, then apply VN forward-pass, and finally take the output $\mathbf{x}^{(I)}$ and {un-standardize} it, i.e.\ apply the inverse of corresponding standardization operation. 
Hereafter the primes are omitted from the above variables for simplicity, although they are implied.

\subsection{Learned potential functions}
For each iteration layer $i$, several potential functions are learned: $\psi$ for the data term and several $\phi_j$ each for a regularization filter. 
These non-linear potential functions are all defined as piece-wise linear functions parameterized by equally spaced knots within a defined interpolation range $[-r, r]$ as in \cite{Vishnevskiy:ReconstructVN}. 
The number of knots is a hyper-parameter of the network and the value of the activation function at each knot is learned during training. 

The range $[-r, r]$ can either be pre-defined (fixed interpolator) or dynamically updated during training (adaptive interpolator). 
For updating such a range over the training with batches $b$, we keep track of a running average $a$ for the maximum value of each (absolute) pre-activation value, computed herein as $a = 0.95 a + 0.05 a_{b}$ where $a_{b}$ is the maximum for the current batch. 
Regularly during training, if the range $r$ is relatively different than the running average $a$, then the range is updated to be closer to $a$.
In this work, we update in particular once every 1000 training batches using the following condition:
\begin{equation}
\mathrm{if} \quad a\not\in[0.95\,r,\; 1.5\,r] \quad \Rightarrow \quad
r \leftarrow 0.7 r + 0.3 a.
\end{equation}

\subsection{VN training}
In order to learn the optimal set of parameters $\mathbf{\Theta} = \{\mathbf{P}^{(i)}, \mathbf{\chi}_d^{(i)}, \mathbf{\psi}^{(i)},  \mathbf{K}_j^{(i)}, \tilde{\mu}_{\mathbf{K}_j^{(i)}}, \mathbf{W}_j^{(i)},\mathbf{\phi}^{(i)}_j,   \mathbf{\chi}_r^{(i)},  \gamma^{(i)} \} $, as the loss function to be minimized during the VN training, we use the $\ell_1$-norm of reconstruction error to the known ground truth $\mathbf{x}^*$ as follows:
\begin{equation}  \label{eq:loss}
     \|\mathbf{\hat{x}} - \mathbf{x}^* \|_1 =  \sum_{p=1}^{P_x} |\mathbf{\hat{x}}_p - \mathbf{x}^*_p|,
\end{equation}
where $P_x$ denotes the number of pixels in the reconstructed SoS image. 
Note that this loss function is applied to the \textit{standardized} SoS images. 

\subsection{Regularization for robustness} 
To make the network robust to changes in input distribution, we herein utilize two regularization techniques.
\subsubsection{Penalizing intermediate network losses}
To regularize training, we use an \textit{exponentially weighted} $\ell_1$-loss~\cite{vishnevskiy:ct} defined as follows:
\begin{equation}
\label{eq:exp_w_loss}
    \mathcal{L}_\mathrm{exp} = \sum_{i=1}^{I} \exp^{-\tau(I-i)}\|\mathbf{x}_i - \mathbf{x}^* \|_1,
\end{equation}
where $\mathbf{x}_i$ is the reconstructed slowness at layer $i$ and $\tau$ is a parameter defining the strength of exponential weighting at intermediate layers. 
When $\tau=0$ all layers of the network are weighted equally which stabilizes the gradients and hence the training of the network.
At the other extreme, $\tau$ approaching $\infty$ corresponds to the intended $\ell_1$-norm loss function in~\eqref{eq:loss}.
Herein, the value of $\tau$ has been fixed to 0.25, chosen during preliminary hyperparameter tuning experiments. 

\subsubsection{Smoothing of learned potential functions}
Similarly to~\cite{hammernik:vnMRI,vishnevskiy:ct}, we herein utilize smooth potential functions.
We ensure smoothness formally by penalizing the $\ell_1$-norm of their second order derivative:
\begin{align}
\label{eq:potf_reg}
\begin{split}
 \mathcal{L}_\mathrm{act}   & = 
 \lambda_d \sum_{i=1}^I \sum_{k=1}^{N_k^{\phi_i}}\sqrt{(y_{k-1}^{\phi_i}- 2y_{k}^{\phi_i}  + y_{k+1}^{\phi_i})^2 + \varepsilon} \\
& + \lambda_{r} \sum_{i=1}^I \sum_{k=1}^{N_k^{\psi_i}}\sqrt{(y_{k-1}^{\psi_i}- 2y_{k}^{\psi_i} + y_{k+1}^{\psi_i})^2 + \varepsilon},
\end{split}
\end{align}
where $\lambda_{d}$ and $\lambda_{r}$ are the penalty terms for $\phi_i$ and $\psi_i$, respectively. $N_k^{(\cdot)}$ is the number of knots parameterizing a potential function, and $y_{k}^{(\cdot)}$ is the value of the corresponding potential function at the $k^\mathrm{th}$ knot. 
$\varepsilon=10^{-6}$ is to avoid division-by-zero in cost gradient computation.

Eventually, for the VN training, the following loss function 
\begin{equation}
\mathcal{L} = \mathcal{L}_\mathrm{exp} + \mathcal{L}_\mathrm{act},
\end{equation}
is optimized with respect to the parameter set $\mathbf{\Theta}$, using Tensorflow backpropagation that computes the necessary gradients.

\section{Multi-source domain training}

\begin{figure}
    \centering
    \includegraphics[width=\linewidth]{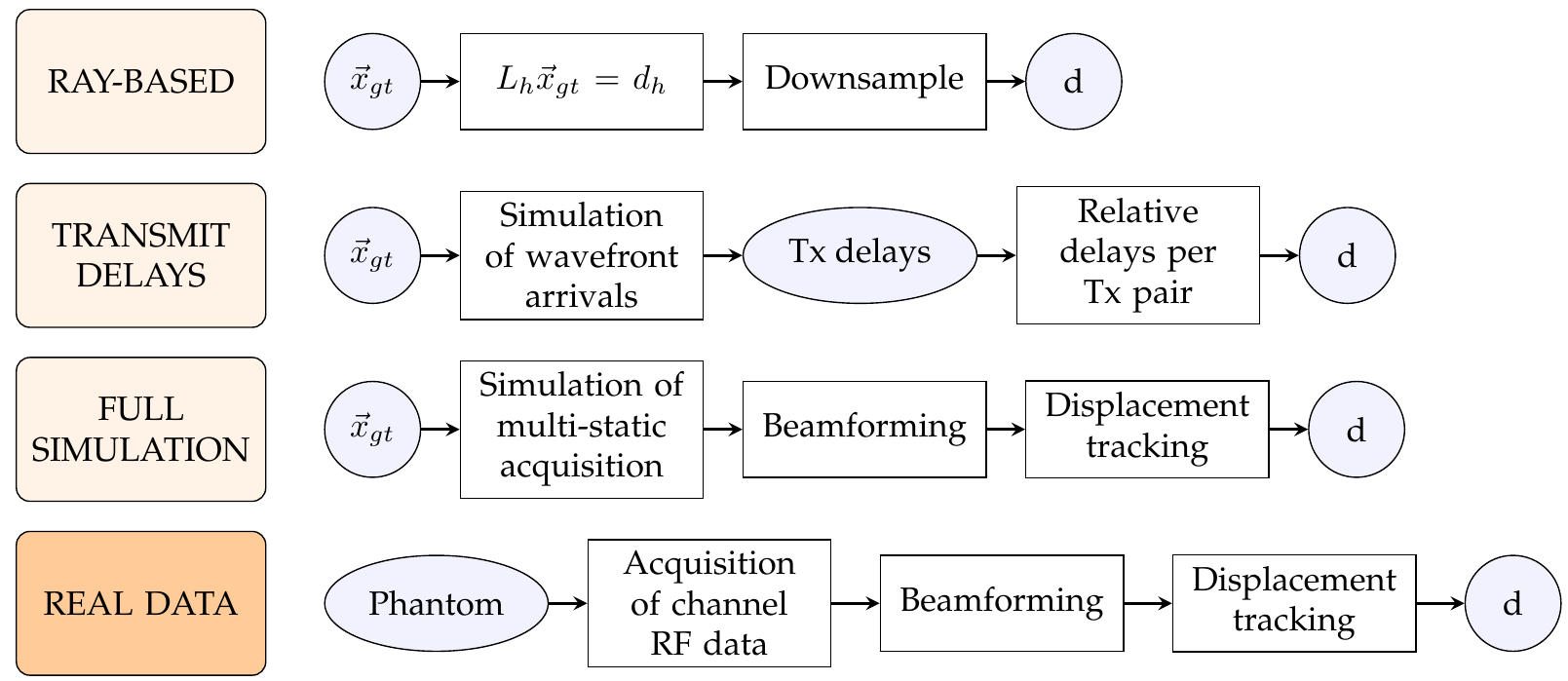}
    \caption{Overview of simple-and-fast to sophisticated-and-slow simulation models, as well as the data acquisition pipeline.}
    \label{fig:fig2}
\end{figure}
In this work we utilize simulated data from several sources, easy and fast to obtain forward problem approximations as well as slow but more representative full wave propagation simulations.
Each simulation method leads to different measurements for the same ground truth SoS map. Fig.~\ref{fig:fig2} summarizes the different data simulation and acquisition procedures described below. 

\subsubsection{Ray-based simulations}
Computationally the most efficient way to simulate training data is a geometric simulation based on the FP. 
Having defined the ground truth SoS map $\mathbf{c}$, one can easily compute measurements using $\mathbf{L}\mathbf{x} = \mathbf{d}$, where $\mathbf{x} = \frac{1}{\mathbf{c}}$ is the inverse SoS map. 
One should however avoid the \textit{inverse crime}~\cite{colton:inversecrime}.
To that end, we use a spatially high resolution SoS map $\mathbf{c}_h$ and corresponding model $\mathbf{L}_h$, to generate high resolution measurements $\mathbf{d}_h$. These measurements were then downsampled to the lower resolution needed for the learned reconstruction, as described in \cite{Vishnevskiy:ReconstructVN}. 

To simulate the missing measurements (so-called undersampling) due to confidence masks, we use two strategies:
(1)~As a simple though not realistic solution, we dismiss some simulated measurements sampled randomly with a fixed \emph{uniform} probability. 
This creates spatially \emph{incoherent} undersampling masks for training. 
With actual acquisitions, however, any missing values from noise are often grouped in ``patches'', since artifacts, wave dispersion, etc occur similarly in nearby regions. 
Hence, (2)~we produce more realistic masks by first generating a low resolution incoherent mask (herein $16 \times 16$ pixels), which is then upsampled to the training grid size (herein $84 \times 64$) using bilinear interpolation. This procedure forms patches, for the measurements to be omitted in the training. 

Moreover, real-world time delay measurements are not perfect due to errors in the beamforming process or in the displacement tracking algorithm. In order to achieve a good performance on real data, the network has to be able to reconstruct the SoS even in presence of noise. 
Therefore, Gaussian noise was also added during training in order to make the network more robust to noisy measurements. In practice, for each data point a noise rate $\eta$ was uniformly sampled between 0 and a maximum noise rate value (chosen at the beginning of the experiment). 
Then, a Gaussian noise is drawn from $\mathcal{N}(0, \eta\, 10^{-7}\, \mathbf{I}_{n_r \times n_r})$, where $\mathbf{I}_{n_r \times n_r}$ is the identity matrix and $n_r$ the size of the input measurement.
This noise is finally added to the input measurement $\mathbf{d}$. 

\subsubsection{Relative transmit delay simulations}
Ray-based measurements are fast to compute. 
Hence, it is easy to obtain a training dataset containing several thousands of data points. 
However, these measurements depend heavily on the assumed ray wave propagation path defined in the $\mathbf{L}$ matrix. 
In order to evaluate the performance of the network in a more realistic context, it is necessary to use more complex wave propagation simulations. The k-Wave Matlab toolbox~\cite{treeby:kwave} allows to perform full wave propagation simulations. 
For a given SoS map, it is thus possible to compute ideal \textit{relative transmit time delays} between any two pairs of Tx channels using these wave propagation simulations. 
By cross-correlating the transmit pulse with the recorded signal at all imaging locations, we estimate the wavefront arrival time to any pixel on the imaging grid.
Then, relative delays between two transmits can be calculated as:
\begin{equation}
\mbox{relativeDelay(Tx1, Tx2)} = \mbox{arrival(Tx1)} - \mbox{arrival(Tx2)}
\end{equation}
where $\text{arrival}(\cdot)$ is a vector containing the arrival times of the wavefront for each pixel in the grid. 
Since we use the same Rx aperture for any Tx as in~\cite{jaeger_computed_2015,Sanabria:spatialreconstsos}, the above \emph{transmit delay} is the expected relative time delay between two frames.
Such \emph{relative delay} simulation thus incorporates the wave nature and corresponding refraction, dispersion, etc.\ effects, meanwhile neglecting the effects and noise from beamforming and displacement estimation.

\subsubsection{Full-pipeline wave simulations} In order to take into account the full-pipeline effects, including beamforming and displacement estimation, we also simulate and record with k-Wave raw channel RF echos at each Rx element.  
The scattering medium is achieved by a heterogeneous density distribution~\cite{rau_sos19}.
Delay-and-sum beamforming and displacement tracking algorithms are then applied to these simulated channel RF data in order to estimate relative time delays. 
Herein we call this as \textit{full-pipeline simulation}, which is computationally intensive but allows to have measurements more representative of the whole processing pipeline similarly to in-vivo or phantom data. 
In particular, these simulated measurements will have undersampled structures, i.e.\ pixels where the displacement estimation algorithm reports low confidence (normalized cross-correlation), which is thus better representative of real acquisition data. 
Note that the quality of the measurements obtained using the full-pipeline simulation is highly dependent on the assumed initial SoS used for the beamforming of each transmit event. This is similar to what is encountered in a physical acquisition scenario and hence represents a more realistic data simulation setting. 
Indeed, if such assumed SoS is very different from the true SoS, then the beamformed images and hence the displacement estimation may have very poor results, with many unreliable and missing measurements. 

\section{Experiments and results}
\subsection{Datasets}
Herein we describe the experimental setup used to train and evaluate reconstruction algorithms. 
Contrary to classic machine learning applications, we want to generalize to both unseen speed-of-sound maps and measurement models, i.e.\ wave propagation settings.
In order to assess results for such a domain shift setup, several evaluation sets were developed, called below a validation set, which follows the distribution of the training set, or a test set, which is outside such distribution including different type of (geometric) inclusions. 

\subsubsection{Training set}
The ground truth SoS maps were designed to contain randomly shaped inclusions, defined as random deformations from random ellipses, as in~\cite{Vishnevskiy:ReconstructVN}. 
Half the inclusions were filtered to yield smooth inclusion edges. 
The background SoS value was also varied smoothly, to make learned networks more robust to such changes as well. 
Finally, we used 5$\%$ of the training set with no inclusions, such that the network is not biased to always output an inclusion. 

\begin{figure}
    \centering
    \includegraphics{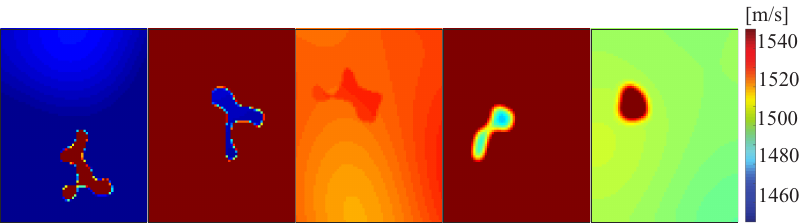}
    \caption{Examples of ground truth SoS maps sampled from custom deformed elliptic inclusions distribution.}
    \label{fig:fig3}
\end{figure}

Figure~\ref{fig:fig3} shows some examples of SoS ground truth maps from the training set, which contains 10,000 images (5,000 sharp and 5,000 smooth inclusions), with an average SoS of 1507\,m/s across all training images.

The main training set was generated with ray-based simulations. 
Indeed, with this simulation method, a very large training set can be acquired quickly. 
Wave-based simulations (both transmit-delays and full-pipeline) are computationally too expensive to acquire a large enough training set. 
Hence, we use only this data as a training baseline.

\subsubsection{Validation set}
The \textit{validation} set was designed to evaluate the robustness of the network against a shift in the measurement model: 
A set of 64 ground truth images was sampled from the training distribution using the full-pipeline simulation. 
The assumed SoS used for beamforming to obtain the full-pipeline data was set to 1510 m/s for all images. 

\subsubsection{Test set}
For the \textit{test set}, we designed custom numerical phantoms containing 28 circular inclusions of various sizes and locations as well as 4 rectangles, all simulated with full-pipeline simulations. 
Images in the test set aim to evaluate reconstruction quality with respect to six different aspects: 1) depth, 2) inclusion size, 3) inclusion edge smoothness, 4) contrast ratio between background and inclusion SoS, 5) level of variation in the background SoS, and 6) for rectangular inclusions, the orientation. 
This test set allows to investigate the behavior of the network in a setting where both the ground truth distribution and the measurement model has changed between training and testing. 
The different sets are summarized in Table~\ref{tab:2}. 
\begin{table}
    \caption{Overview of the different datasets used in the experiments. 
    Number of generated data points for each dataset is given in parentheses. }
    \label{tab:2}
    \centering
    \begin{tabular}{@{}>{\raggedright}p{0.35\linewidth}ll}
    \toprule
    Dataset \\ Simulation approach & Ray-based & Full pipeline \\
   \midrule
       Random inclusions  & Training set ($10^4$) & Validation set (64) \\
       Geometric primitives & --- & Test set (32) \\
       \bottomrule
    \end{tabular}
\end{table}

The reconstruction quality was assessed in terms of Root Mean Squared Error (RMSE) to the ground truth SoS (when available).  
The RMSE is defined as:
\begin{equation}
    \text{RMSE}(\mathbf{x}^*, \mathbf{\hat{x}}) = \sqrt{\frac{1}{P_x}\sum_{p=1}^{P_x}\left(\mathbf{x}^*_p- \mathbf{\hat{x}}_p\right)^2}
\end{equation} 
where $P_x$ is the number of pixels in the image, $\mathbf{x}^*$ designates the ground truth and $\mathbf{\hat{x}}$ the reconstructed SoS map.  

\subsubsection{Phantom data}
Images from a breast phantom (CIRS Multi-Modality Breast Biopsy and Sonographic Trainer, Model 073, CIRS Inc., Norfolk, VA, USA) was acquired using a UF-760AG ultrasound system (Fukuda Denshi, Tokyo, Japan) with FUT-LA385-12P linear array transducer ($N_c = 128$ channels, $300\,\mu\mathrm{m}$ pitch, $f_c=5\,\mathrm{MHz}$ center frequency, 4 half cycles pulses). 
The phantom mimics the tissue material and the geometry of the breast and contains both hard and cystic inclusions. 
For phantom data, B-mode images help to locate the inclusions. 
However, given the absence of ground truth SoS map, on this dataset the quality of the reconstruction could only be assessed qualitatively or in terms of contrast to noise (CNR) ratio defined as \cite{Patterson:CNR}:
\begin{equation}
    \frac{|\mu_{inc} - \mu_{bg}|}{\sqrt{\sigma_{inc}^2+\sigma_{bg}^2}}, \end{equation}
where $\mu_{inc}$ and $\mu_{bg}$ are the mean SoS in the inclusion and the background, respectively, and similarly $\sigma_{inc}$ and $\sigma_{bg}$) are the corresponding standard deviations. 
The inclusions were manually delineated based on the B-mode image. 

\subsection{Numerical simulation experiments}
VN experiments were implemented in Python using Tensorflow~\cite{abadi2016tensorflow}. 
The training objective was minimized through mini-batch optimization with the ADAM optimizer~\cite{Kingma:adam} with a batch size of 16 images and 120,000 training iterations. 
The measurements were undersampled with an undersampling rate between 10\% and 90\% and a noise rate of up to 10\%. 
\subsubsection{Determining model complexity} 
\begin{figure}
    \centering
    \includegraphics[width=\linewidth]{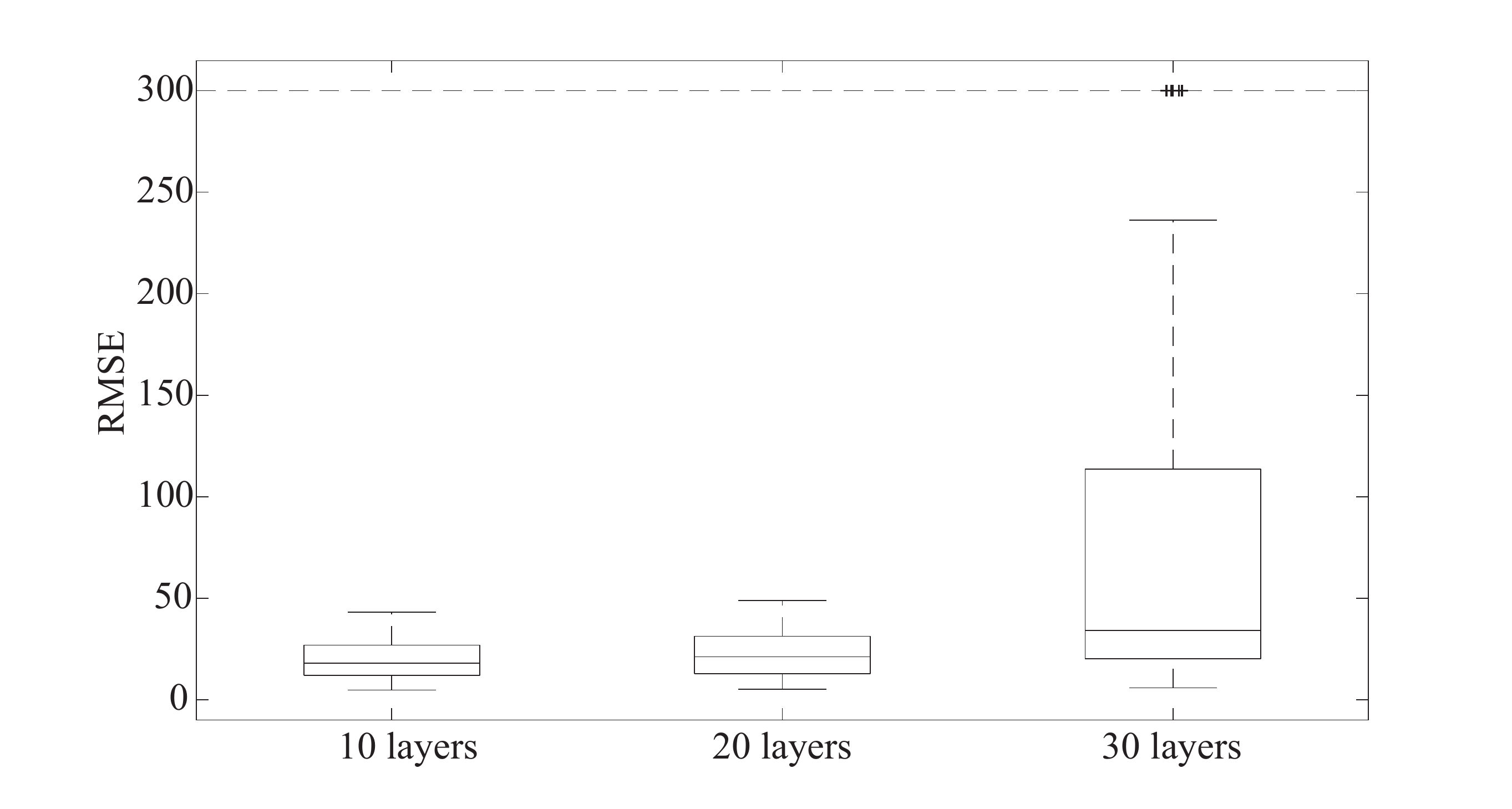}
    \caption{RMSE results on the fullpipeline validation set for models with 10, 20, and 30 layers, having all other parameters equal.
    Outliers thresholded at 300, while some outliers for 30 layers have RMSE up to 2000.
    On each box, the central mark indicates the median, and the bottom and top edges are the $q_b$ 25$^\text{th}$ and $q_t$ 75$^\text{th}$ percentiles, respectively. The whiskers are then defined as $q_b - 1.5(q_t-q_b)$ and $q_t + 1.5(q_t-q_b)$. Outliers are represented by '+' symbols.}
    \label{fig:9}
\end{figure}

One essential parameter of the VN architecture  is the number of unrolled iterations to perform, i.e.\ the depth of the network.
Increasing the number of layers typically increases the complexity of the model. 
This enables the network to learn a more complex data manifold. 
However, it may also lead to overfitting to the training set. 
Hence, there is a trade-off between having a model that is sufficiently complex to learn the transformation from measurements to SoS maps, but still keeping a good generalization capability. 
Figure~\ref{fig:9} shows the results of the models trained above on the full-pipeline validation set, which indicates that the model with 30 layers is less robust to such acquisition model domain shift, compared to the 10- and 20-layer models. 
This complex model seems to learn properties (including potential artifacts) specific to the ray-based measurements on which it was trained. 
This experiment not only helps us determine a network depth, but also highlights the necessity to evaluate the network on a full-pipeline validation set, as results on the ray-based training set may be misleading by hiding an overfitting pattern.  

\subsubsection{Evaluating the domain shift}
We evaluate different versions of VN with our proposed contributions, especially in terms of generalizability and robustness to domain shift. 
Five following models are compared:
\begin{itemize}
    \item \textit{Raw}: a standard VN with 20 layers, 32 filters as presented in Algorithm~\ref{algo:finalSoSVN}, using a standard $\ell_1$-loss function, trained on ray-based data only.
    \item \textit{ExpNoSmooth}: similar to Raw, but using an exponentially weighted loss (\ref{eq:exp_w_loss}) with $\tau = 0.25$.
    \item \textit{ExpSmooth}: similar to ExpNoSmooth, but additionally adding activation function smoothness penalty~(\ref{eq:potf_reg}) for the regularization term using a regularization constant of $\lambda_r=10^5$, chosen empirically. 
    \item \textit{ExpSmoothMixed}: similar to ExpSmooth, but trained on a mix of ray-based data, relative transmit delays, and full-pipeline simulations. During training, the mixing was achieved by sampling 11 images from the ray-based training set (of 10,000), 4 images from the full-pipeline training set (of 870), and 1 image from the relative transmit time delays (of 630) for each mini-batch. The mixing proportion has been investigated and optimized with preliminary experiments.
    \item \textit{L-BFGS}: our state-of-the art baseline, as an iterative optimization of the regularized AWTV objective~\cite{rau_sos19}. 
\end{itemize}

\begin{figure}
    \centering
\begin{subfigure}[b]{0.23\columnwidth}
    \includegraphics[width=\columnwidth]{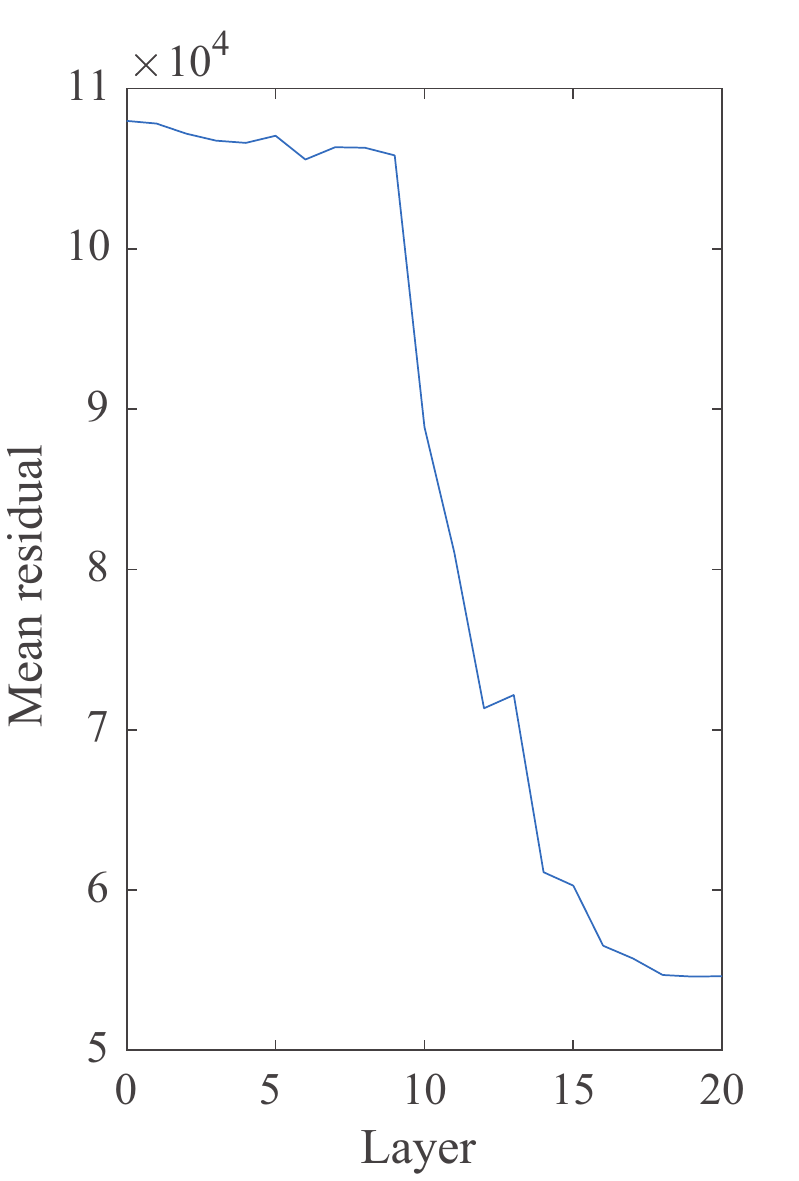}
    \caption{}
    \label{fig:8_a}
\end{subfigure}
\hfill
\begin{subfigure}[b]{0.75\columnwidth}
    \includegraphics[width=\columnwidth]{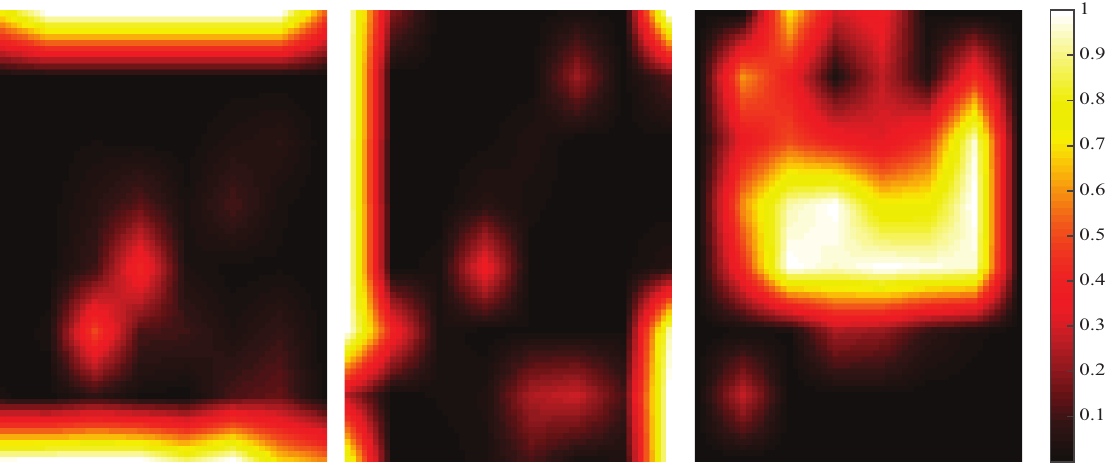}
    \caption{}
    \label{fig:8_b}
\end{subfigure}
\caption{(a) Visualization of the reconstruction error per layer on the ray-based validation set \textit{ExpSmoothMixed} model. (b)~Examples of learned spatial weights.}
\end{figure}

The introduction of the regularization techniques particularly stabilize the unrolled behavior learned by the network, with a regular decrease of the error along the layers, demonstrated in Fig.~\ref{fig:8_a} for the ray-based validation set.
In Fig.~\ref{fig:8_b} we show sample learned spatial weights $\mathbf{W}_j^{(i)}$, illustrating some regularizer terms specializing on image center while some others on certain edges. 

\begin{figure}
\centering
\begin{subfigure}[t]{\columnwidth}
\centering
    \includegraphics[width=.8\columnwidth]{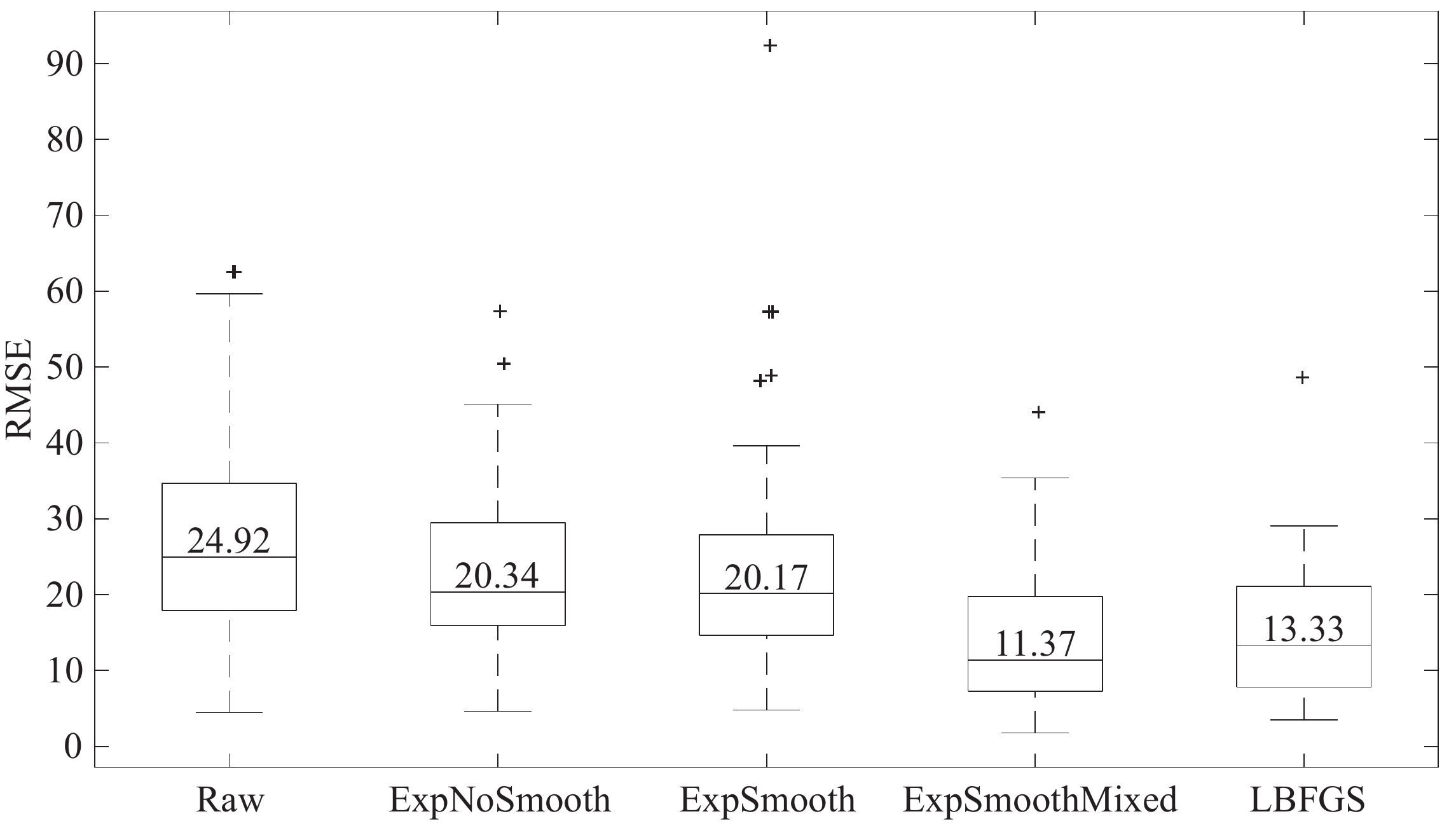}
    \caption{Validation set with deformed elliptic inclusions}
    \label{fig:7_a}
\end{subfigure} \\
\begin{subfigure}[t]{\columnwidth}
\centering
    \includegraphics[width=.8\columnwidth]{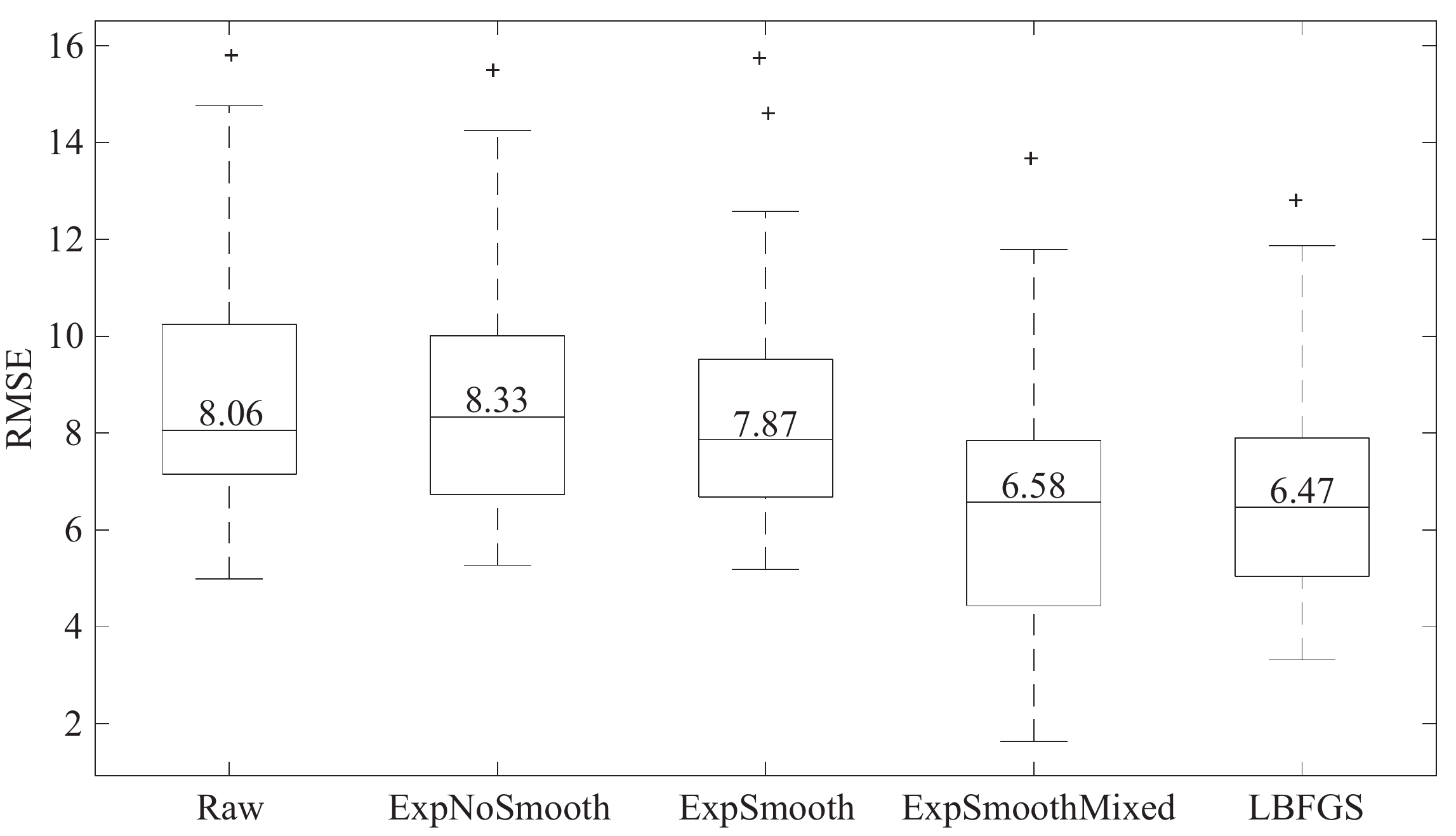}
    \caption{Test set with circular and rectangular inclusions}
    \label{fig:7_b}
\end{subfigure}
\caption{RMSE results on full-pipeline validation and test sets for various VN models compared to L-BFGS: \textit{Raw} is the initial VN model, trained on ray-based data only; \textit{ExpNoSmooth} is the VN with exponential weighting, without activation function smoothing, trained on ray-based data only; \textit{ExpSmooth} adds activation function smoothing; \textit{ExpSmoothMixed} is additionally trained on a mix of ray-based, relative transmit delays, and full-pipeline data.}
\label{fig:7}
\end{figure}

\begin{figure*}
    \hspace{0.51ex}\includegraphics[width=0.99\linewidth]{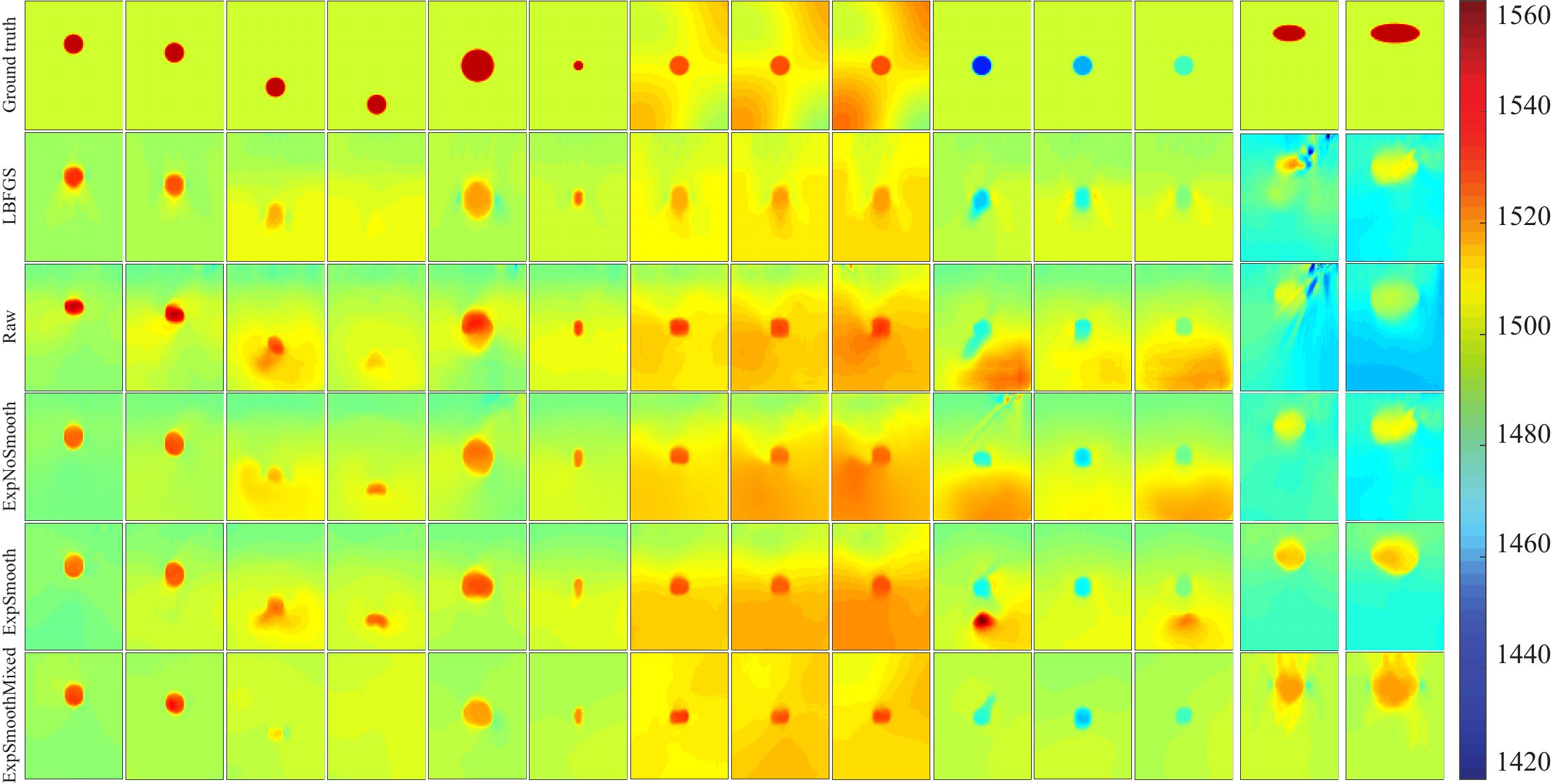}\\
    \includegraphics[width=.923\linewidth]{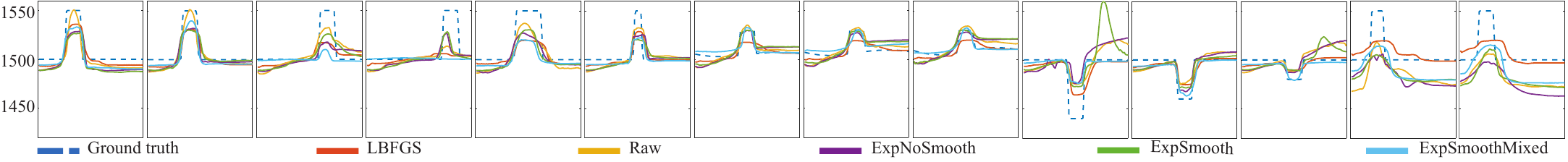}
 \caption{Sample reconstructions from the test set for different VN models in comparison to the groundtruth and the L-BFGS baseline. Bottom row shows vertical cross-sections along the centerline of SoS reconstructions.
 }
\label{fig:5}
\end{figure*}

In Fig.~\ref{fig:7} we can see that the introduction of the exponentially weighted loss function together with activation function smoothing improves the results on both full-pipeline validation and test sets.
Nevertheless, the main source of improvement is seen to originate from the multi-domain source training. 
\textit{ExpSmoothMixed} improves the median RMSE by 54\% on the validation set with the full-pipeline dataset, compared to the baseline \textit{Raw} model. 
The test set results (with image domain shift as well) in Fig.~\ref{fig:7} corroborate these promising findings, showing the superiority and robustness of \textit{ExpSmoothMixed} compared to its alternatives.
Indeed, \textit{ExpSmoothMixed} is the only VN model reaching the L-BFGS performance (and even surpassing it in the validation dataset). 
For illustration purposes, the reconstruction results on a subset of images of the test set are depicted in Figure~\ref{fig:5}. 
We can see that the reconstruction quality is sensitive to inclusion location, in particular its depth; i.e.\ the deeper the inclusion is, the worse its reconstruction. 
This is due the increased imprecision of displacements estimates at deeper locations and the less number of rays/measurements contributing to the estimation of deeper pixels.
Furthermore, the SoS contrast is sometimes underestimated, potentially due to spatial regularization effects. 
Note that any systematic bias may also originate from prior SoS assumption in beamforming~\cite{rau_aberration_2019} causing errors, e.g. in displacement tracking.
Notwithstanding, we can see that overall the inclusions are reconstructed most accurately by our proposed \textit{ExpSmoothMixed} model.

\subsection{Results on phantom data}
\begin{figure*}
    \centering
    \includegraphics[width=\linewidth]{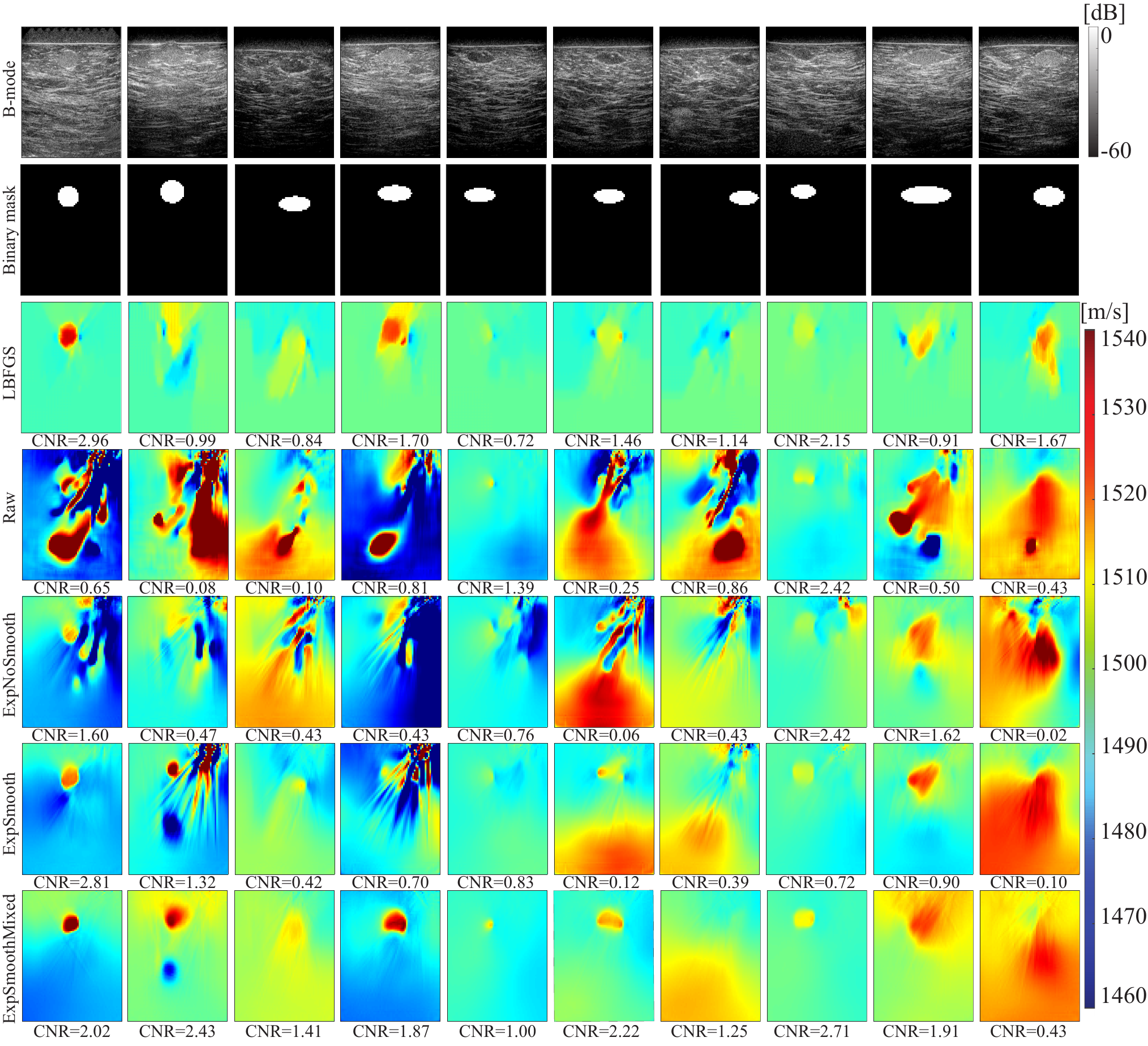}
    \caption{Example SoS reconstructions from breast phantom data using different VN models in comparison to L-BFGS.  Physical inclusions have varying stiffness, without a groundtruth measurement, but the B-mode visible inclusions were marked as shown in second row for reporting reconstruction CNR.}
    \label{fig:fig6}
\end{figure*}

As can be seen from Figure~\ref{fig:fig6} the breast phantom reconstructions contain pronounced striking artefacts for VNs that are trained with the ray-based model.
In terms of CNR, the \textit{ExpSmoothMixed} VN model qualitatively outperforms \textit{L-BFGS} for 8 out of 10 experiments and visually provides the most coherent SoS maps.
Typical reconstruction time with \textit{ExpSmoothMixed} was 12\,ms on NVIDIA Titan V GPU and 48\,s with \textit{L-BFGS} on a 4-core Intel CPU.

\section{Discussion and Conclusions}
We presented herein a framework for real-time US SoS imaging based on the variational image reconstruction neural networks. 
The novel approach for network learning and regularization allows us to leverage synthetic training data for accurate and robust reconstruction of breast phantom acquisitions.
Our first regularization approach is the exponential weighting of the loss function. 
By including the reconstruction error of the intermediate layers in the loss function, the network is encouraged to steadily decrease an error at each consecutive layer. 
Without this weighting, the intermediate reconstruction results can be unstable and contain little information about the estimated SoS image, cf.\ compare \textit{ExpNoSmooth} vs.\ \textit{Raw}.
Although, without exponential weighting, the network is less constrained to achieve final output reconstructions, such fewer constraints appears to lead to less robustness in presence of domain shift. 
Note that shifts in the image domain may lead to different filter response distributions.
Therefore, adding our second proposed regularization approach as the smoothing of activation functions is particularly important and beneficial, cf. compare \textit{ExpSmooth} vs.\,\textit{ExpNoSmooth}.

Despite the above adaptations, our VN trained using the traditional single-domain ray-based training, e.g.\ \emph{ExpSmooth}, performed consistently inferior to the L-BFGS baseline. 
This is true both for the acquired phantom datasets as well as the full-pipeline simulations, indicating an significant domain-shift already from ray-based to full-pipeline data.
Since the full-pipeline simulation is the most representative of actual acquisitions, we used full-pipeline validation data as a proxy to tune our model and training procedure.
This also hints at the potential of improved generalization ability by incorporating full-pipeline data as well as potentially other domains in the training, which led us to our proposed multi-domain training procedure.
Our experiments demonstrate that leveraging different training data sources yields a more generalizable network, enabling more accurate reconstructions on the phantom data.
Compared to the baseline unregularized \textit{Raw} 20-layer VN, the proposed \textit{ExpSmoothMixed} VN allowed to reduce median RMSE by 54\% on the full-pipeline validation set. 
Our proposed model \textit{ExpSmoothMixed} was the only model capable of reaching the L-BFGS accuracy on this dataset. The results on the full-pipeline test set moreover confirm these findings in a setting where both the measurement as well as the ground truth distributions are shifted.
Note that the validation set yields higher errors than the test set, due mainly to the former having more complex inclusions and background variation as well as a lower inclusion contrast (on average, 0.5\% for the validation vs.\ 2.8\% for test set).

Qualitative and CNR evaluation on phantom data confirm significant improvements from using  \textit{ExpSmoothMixed}. 
On our phantom evaluation, the proposed robust model \textit{ExpSmoothMixed} even outperformed the iterative L-BFGS solution in terms of CNR for 8 images out of 10.
Similar improvements are observed in Fig.~\ref{fig:5} with the full-pipeline dataset.
This shows that training on various data sources and adding suitable regularization terms to the network can help avoid overfitting to the training data manifold, by constraining the network to learn features that are more domain invariant. 

Similarly to many other studies \cite{Krueger98,Huang2004, Sanabria:SoSReflector,duric_detection_2007,gemmeke_3d_2007,malik_breast_2019,Sanabria:spatialreconstsos, jaeger_computed_2015,stahli_forward_2019,Vishnevskiy:ReconstructVN,rau_sos19,LisaRuby-InvestigativeRadiology-Jul2019}, we herein utilize straight ray assumption for wavefront propagation in tissue although extensions of our proposed methods to bent ray approximations as in \cite{Li2009,Li2010,Jirik12} can be envisioned.
We herein do not employ adaptive receive aperture~\cite{stahli_forward_2019}, to keep the utilized simulation modalities more consistent since transmit delay simulations cannot incorporate differences in receive paths.

With the proposed method, we enable real-time SoS imaging based on inverse-problems with learned regularizers and activation functions, which would otherwise need computationally expensive iterative solvers.
By reducing the inference times to 12\,ms per image, we here pave the way for SoS image applications on a large scale, as the proposed VN is implementable with conventional and widely available US systems and transducers.
With fast and accurate SoS reconstruction, we furthermore open up the possibility for aberration corrections, which is important for numerous other ultrasound imaging modalities. 
This is for instance studied in~\cite{jaeger_full_2015,rau_aberration_2019}, where B-Mode image resolution is seen to be improved by correcting the aberration effects introduced by SoS heterogeneities. 

Our method can be further extended to reconstruct other wave propagation characteristics, such as attenuation. 
With conventional US system, attenuation imaging can be achieved as in~\cite{rau_attenuation_2019} using an iterative solution to an inverse problem. 
Accordingly, our proposed method can be translated for this and other inverse problem based image reconstruction problems, improving surely the computational times as well potentially the accuracy of reconstructions.
We use a subgradient approach in our VN.  
Although proximal methods for iterative image reconstruction tend to find sparser solutions, such algorithms are less efficient for sparsity models in the \emph{analysis form}~\cite{Fessler20}, which we employ. 
Unrolling of variable splitting methods such as ADMM, which was presented for certain structures of imaging matrices, e.g. Cartesian Fourier transform in~\cite{Yang16}, can be adapted to accelerate convergence in our imaging problem.
Besides the multi-domain training as shown herein, in future work domain adaptation techniques such as domain adversarial training~\cite{Ganin16,Kamnitsas17} could be investigated and further improve tomographic reconstructions.

\bibliographystyle{IEEEtran}
\bibliography{references}

\end{document}